\documentclass[conference]{IEEEtran}
\IEEEoverridecommandlockouts
\usepackage{cite}
\usepackage{acronym}
\usepackage{makecell}
\usepackage[citecolor=black,
            colorlinks=true,
            linkcolor=black,
            urlcolor  = black,
            pdftitle={},
            pdfauthor={},
            pdfkeywords={},
            pdfsubject={}]{hyperref}
\usepackage[english]{babel}
\addto\extrasenglish{}
\addto\extrasenglish{}
\addto\extrasenglish{}
\addto\extrasenglish{}
\addto\extrasenglish{}
\addto\extrasenglish{}
\usepackage{amsmath,amssymb,amsfonts}
\usepackage{flushend} 
\usepackage{algorithmic}
\usepackage{graphicx}
\usepackage{pgf}
\usepackage{textcomp}
\usepackage{xcolor}
\usepackage{color}
\usepackage{colortbl}
\usepackage{booktabs}
\usepackage{svg}
\usepackage{multirow}
\def\BibTeX{{\rm B\kern-.05em{\sc i\kern-.025em b}\kern-.08em
    T\kern-.1667em\lower.7ex\hbox{E}\kern-.125emX}}

\acrodef{ASR}       {automatic speech recognition}
\acrodef{CEC1}      {1$^{st}$ Clarity Enhancement Challenge}
\acrodef{CEC2}      {$2^{nd}$ Clarity Enhancement Challenge}
\acrodef{CNN}       {convolutional neural network}
\acrodef{CPC2}      {2$^{nd}$ Clarity Prediction Challenge}
\acrodef{HA}        {hearing aid}
\acrodef{FIR}       {finite-impulse response}
\acrodef{HASPI}     {hearing aid speech perception index}
\acrodef{HSI}       {human speech intelligibility}
\acrodef{MSE}       {mean-squared error}
\acrodef{PESQ}      {perceptual evaluation of speech quality}
\acrodef{SI}        {speech intelligibility}
\acrodef{SI-SNR}    {scale-invariant signal-to-noise ratio}
\acrodef{SNR}       {signal-to-noise ratio}
\acrodef{SSSR}      {self-supervised speech representation}
\acrodef{STOI}      {short-term objective intelligibility}
\acrodef{SQ}        {speech quality}

\begin{document}

\title{Using Speech Foundational Models in Loss Functions for Hearing Aid Speech Enhancement
\thanks{This work was supported by the Centre for Doctoral Training in Speech and Language Technologies (SLT) and their Applications funded by UK Research and Innovation [EP/S023062/1] and EPSRC Clarity Project [EP/S031448/1]. This work was also supported by WS Audiology and Toshiba.}
}

\author{\IEEEauthorblockN{Robert Sutherland, George Close, Thomas Hain, Stefan Goetze, and Jon Barker}
\IEEEauthorblockA{\textit{Speech and Hearing (SPandH) group, Department of Computer Science}, 
\textit{The University of Sheffield}, 
Sheffield, UK \\
\{rwhsutherland1, glclose1, t.hain, s.goetze, j.p.barker\}@sheffield.ac.uk}
}

\maketitle

\begin{abstract}
Machine learning techniques are an active area of research for speech enhancement for hearing aids, with one particular focus on improving the intelligibility of a noisy speech signal. Recent work has shown that feature encodings from self-supervised speech representation models can effectively capture speech intelligibility. In this work, it is shown that the distance between self-supervised speech representations of clean and noisy speech correlates more strongly with human intelligibility ratings than other signal-based metrics. Experiments show that training a speech enhancement model using this distance as part of a loss function improves the performance over using an SNR-based loss function, demonstrated by an increase in HASPI, STOI, PESQ and SI-SNR scores. This method takes inference of a high parameter count model only at training time, meaning the speech enhancement model can remain smaller, as is required for hearing aids.
\end{abstract}


\section{Introduction}

Hearing impairment is a widespread problem worldwide~\cite{davis2019hearing} and especially in countries with ageing populations~\cite{renniesHI4aging,park2020population}. There is an emerging interest in the development of hearing aid systems which incorporate neural network components~\cite{graetzerClarity2021ChallengesMachine2021,cornell2022_clarity,diehl_2023,ouyang2023,schroter2022,fedorov20_interspeech}. The particular focus of this development has been on increasing the \textit{intelligibility} of speech in the processed input audio. A core challenge in this area is that neural hearing aid models must be computationally efficient to fit within hardware constraints. 
Standard loss functions used to train neural network-based hearing aid systems operate at a signal level, e.g., based on \ac{SNR}, \ac{SI-SNR}, or \ac{STOI}~\cite{Braun21LossFunctions,leRoux19_SISDR,taalAlgorithmIntelligibilityPrediction2011}. However, recent work has shown that representations sourced from large speech foundational models (i.e., with high parameter count), such as Whisper \cite{radfordRobustSpeechRecognition2022a} or WavLM \cite{chenWavLMLargeScaleSelfSupervised2022}, are able to well capture information about the intelligibility of the signal \cite{close2023non,cuervoSpeechFoundationModels2024,mogridge2024nonintrusive}. Previous work has explored how the distances between the \ac{SSSR} representations of clean speech and noisy speech correlate with speech quality metrics and how they can be used in loss functions \cite{closePerceivePredictSelfsupervised2023} for single-channel noise reduction. However, the correlation with speech intelligibility metrics and application of \ac{SSSR} based loss functions for hearing aid systems have not yet been well explored in the existing literature.  
 
This paper investigates how distances between WavLM representations correlate with \ac{SI} and \ac{SQ} metrics like the \ac{HASPI}~\cite{katesHearingAidSpeechPerception2021}, \ac{STOI}~\cite{taalAlgorithmIntelligibilityPrediction2011}, \ac{PESQ}~\cite{pesq01ICASSP}, and others, as well with real-world \ac{HSI} ratings from listening tests, released as part of the \ac{CPC2} \cite{barker24_icassp}. Further, the distances of the WavLM representations are evaluated for their usefulness as loss functions to train the denoising component of a neural hearing aid system. The proposed method shows an increase in terms of perceptually motivated enhancement metrics versus a traditional baseline~\cite{tuTwoStageEndtoEndSystem}, while maintaining the strict model parameter count, speed and memory consumption constraints typical of a hearing aid system. This is achieved by taking inference of the much larger WavLM model only during model training.

The structure of the paper is as follows. \autoref{sec:wavlm} introduces the use of WavLM in this work and provides some background of the prior work which uses \acp{SSSR} for speech enhancement. \autoref{sec:analysis} defines a distance measure using the WavLM representations of signals, and investigates how this correlates with human intelligibility labels and perception-based metrics. \autoref{sec:system} describes the system which is subsequently used for the signal enhancement experiments in \autoref{sec:experiment}, using the WavLM-based distance metric as a loss function. Results are presented in \autoref{sec:results} and \autoref{sec:conclusion} concludes the paper.

\section{Self-Supervised Speech Representations for Speech Enhancement}
\label{sec:wavlm}

\Acp{SSSR} such as wav2vec \cite{schneider2019wav2vec} and HuBERT \cite{hsu2021hubert} were initially designed and evaluated as front-end feature extractors for down-stream \ac{ASR} tasks. More recently, WavLM was developed to support a broader set of speech processing tasks, including, e.g., speaker verification, or speaker diarization, which it achieves by not only learning masked speech prediction, but also by learning speech denoising during training \cite{chenWavLMLargeScaleSelfSupervised2022}.

Recent work has shown that WavLM and other \acp{SSSR} are powerful feature representations for quality~\cite{close2023multicmgan} and intelligibility prediction~\cite{close2023non}. Furthermore, the ability of \ac{SSSR}-derived representations to encode quality and intelligibility-related information has been exploited in loss functions for the training of speech enhancement systems~\cite{closePerceivePredictSelfsupervised2023,close2023effect}, showing improved performance in comparison to traditional loss functions for speech enhancement. 
\begin{figure}[!ht]
 \centering
 \resizebox{1.2\columnwidth}{!}{%
 \graphicspath{{figs}} 
 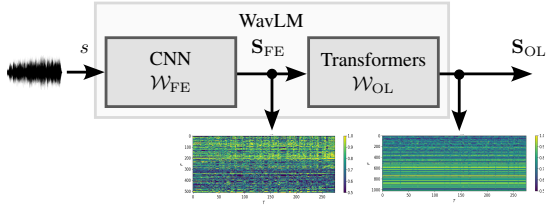 
 }
 \caption{Representations extracted from WavLM model stages.}
\label{fig:wavlm_sages}
\end{figure}
WavLM is a neural network with two distinct stages as shown in \autoref{fig:wavlm_sages}: a \ac{CNN} based encoder $\mathcal{W}_\mathrm{FE}$ and a Transformer~\cite{transformer} based output stage $\mathcal{W}_\mathrm{OL}$. The input to the encoder $\mathcal{W}_\mathrm{FE}$ is a time-domain signal $s$ and its output representation $\mathbf{S}_\mathrm{FE}$ has a dimension $F_\mathrm{FE} \times T$ with encoder output feature dimension $F_\mathrm{FE}=512$ for WavLM~\cite{chenWavLMLargeScaleSelfSupervised2022} and $T$ representing the variable number of frames depending on the length of the input audio $s$. $\mathbf{S}_\mathrm{FE}$ is the input to the Transformer-based WavLM decoder stage $\mathcal{W}_\mathrm{OL}$ which returns a final output representation $\mathbf{S}_\mathrm{OL}$ with dimensions $F_\mathrm{OL} \times T$. For WavLM the output layer feature dimension is $F_\mathrm{OL}=768$~\cite{chenWavLMLargeScaleSelfSupervised2022}. Based on findings from previous work~\cite{closePerceivePredictSelfsupervised2023,close2023effect}, only the encoder representation $\mathbf{S}_\mathrm{FE}$ is used in the following. In this work, the WavLM Base\footnote{\url{https://huggingface.co/microsoft/wavlm-base}} model trained on the Librispeech $960$ hour dataset~\cite{librispeech} is used, which has $94.7$ million parameters. 

\section{WavLM Representation Distances}
\label{sec:analysis}

Initially, the distance between WavLM encoder representations of clean and enhanced speech signals is explored, with a loss function defined as
\begin{equation}
    \label{eq:EncoderOutputDistance}
    \hspace{-1.5ex}\mathcal{L}_{\mathrm{WLM}}\left(s, \hat{s}\right) \hspace{-0.3ex}=\hspace{-0.3ex} \frac{1}{TF_\mathrm{FE}}\sum^T_{t=1}\sum^{F_\mathrm{FE}}_{f=1}\left(\mathbf{S}_\mathrm{FE}[t,f] - \mathbf{\hat{S}}_\mathrm{FE}[t,f]\right)^2\hspace{-1ex},
\end{equation}
where $\mathbf{S}_\mathrm{FE}=\mathcal{W}_\mathrm{FE}(s)$ and $\mathbf{\hat{S}}_\mathrm{FE}=\mathcal{W}_\mathrm{FE}(\hat{s})$ are the representations of the target and estimated signal, respectively. Please note that frame (time) index $t$ and feature index $f$ are omitted for brevity at most places in this paper.

This distance measure is compared with the \acf{SNR} based loss function
\begin{equation}
    \label{eqn:SNR-loss}
    \mathcal{L}_{\mathrm{SNR}}\left(s, \hat{s}\right) = -10\log_{10}\left( \frac{\Vert s \Vert ^2}{\Vert s - \hat{s} \Vert ^2 + \tau\Vert s \Vert ^2}\right),
\end{equation}
which is commonly used for training the speech enhancement component of hearing aid systems, with $\tau=10^{-\mathrm{SNR}_{\mathrm{max}}/10}=0.001$ (where $\mathrm{SNR}_{\mathrm{max}}=30\text{dB}$) to prevent sufficiently denoised signals from dominating the training, as described in \cite{wisdomUnsupervisedSoundSeparation2020}.

In the following, an analysis of how $\mathcal{L}_\mathrm{SNR}$ and $\mathcal{L}_\mathrm{WLM}$ correlate with human intelligibility labels, as well as other perception-based metrics will be conducted.

\subsection{The \texorpdfstring{$2^{nd}$}{2nd} Clarity Prediction Challenge (CPC2) Dataset}

The \ac{CPC2} dataset is used to investigate the relationship between $\mathcal{L}_\mathrm{WLM}$ in (\ref{eq:EncoderOutputDistance}) and $\mathcal{L}_\mathrm{SNR}$ in (\ref{eqn:SNR-loss}), and existing speech metrics and \ac{HSI} labels. The task of \ac{CPC2} is to predict the intelligibility of a signal for listeners with hearing impairment.

\begin{figure}[!h]
    \centering
    \input{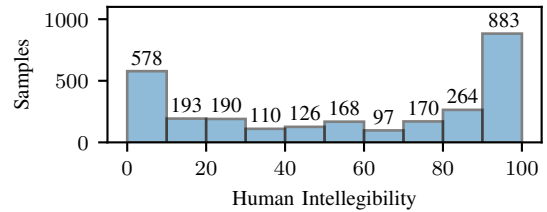}
    \caption{Distribution of human intelligibility labels in the \ac{CPC2} train set}
    \label{fig:cpc2-dist}
\end{figure}

The main signal assessed in \ac{CPC2} is the binaural output signal for left and right channels $\mathbf{\hat{s}} = [\hat{s}_{l}, \hat{s}_{r}]=\mathcal{M}\left(\mathbf{x}\right)$, processed by a \ac{HA} $\mathcal{M}(\cdot)$ with $6$-channel noisy input signal $\mathbf{x}$ (for $3$ microphones in each (left and right) \ac{HA}), containing a clean speech signal $\mathbf{s}$. The samples in the \ac{CPC2} dataset include an enhanced signal $\mathbf{\hat{s}}$, a target signal $\mathbf{s}$, and a \ac{HSI} label, which is the proportion of words that a hearing-impaired listener was able to identify in $\mathbf{\hat{s}}$, as well as the audiogram of the respective listener~\cite{barker24_icassp}. The first training split is used, which contains $2779$ samples; the distribution of these samples is shown in \autoref{fig:cpc2-dist}. For each sample in one of the training splits, $\mathcal{L}_\mathrm{WLM}$ and $\mathcal{L}_\mathrm{SNR}$ were calculated as in (\ref{eq:EncoderOutputDistance}) and (\ref{eqn:SNR-loss}), respectively.

\subsection{Correlation Analysis}
\label{ssec:CorrelationAnalysis}

\begin{figure*}
    \centering
    \resizebox{0.9\textwidth}{!}{%
    \input{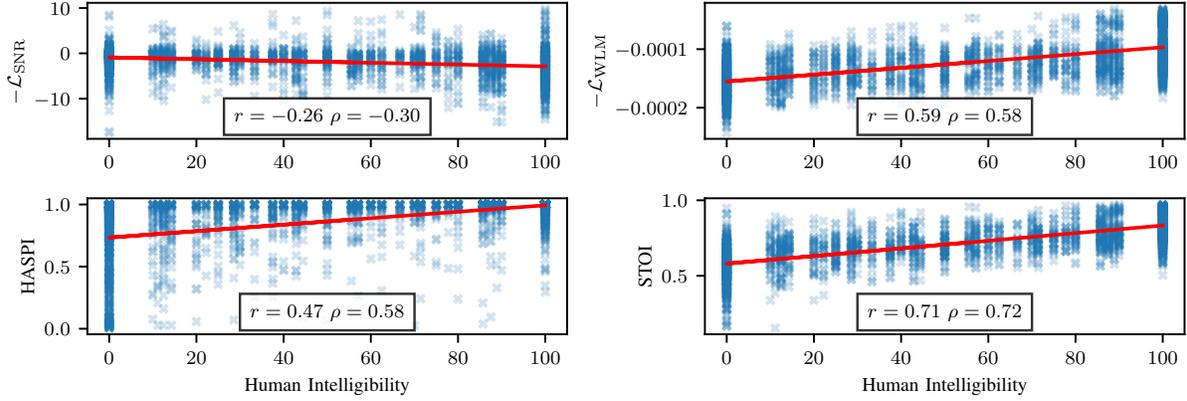}
    }
    \caption{Correlations of $-\mathcal{L}_\mathrm{SNR}$, $-\mathcal{L}_\mathrm{WLM}$, \ac{HASPI} and \ac{STOI} with the human intelligibility labels for the \ac{CPC2} dataset.}
    \label{fig:correlations}
\end{figure*}

\autoref{table:correlations} shows Pearson correlations between \ac{HSI} labels, $-\mathcal{L}_\mathrm{SNR}$, $-\mathcal{L}_\mathrm{WLM}$, \ac{SI-SNR}, \ac{HASPI}, \ac{STOI}, and \ac{PESQ}, with scatter-plots visualising some of the correlations with human intelligibility labels in \autoref{fig:correlations}. Negative values of $\mathcal{L}_\mathrm{SNR}$ and $\mathcal{L}_\mathrm{WLM}$ are used so that positive correlations are expected for all rows and columns. For each objective metric, values for each (left and right) channel are computed and the better score is taken.

\begin{table}[!h]
    \centering
    \caption{Pearson correlations $r$ between \ac{HSI}, losses under test and various speech metrics on the \ac{CPC2} dataset.}
    \resizebox{\columnwidth}{!}{%
    \begin{footnotesize}
    \begin{tabular}{c|ccccccc}
    \toprule
        {}    &{}             &\multicolumn{2}{c}{Losses}&SI-         & {}          &{}           &{}           \\
        \multirow{-2}{*}{Scores}    &\multirow{-2}{*}{\ac{HSI}}             &-$\mathcal{L}_{\mathrm{WLM}}$&-$\mathcal{L}_{\mathrm{SNR}}$&SNR         &\multirow{-2}{*}{HASPI}          &\multirow{-2}{*}{STOI}           &\multirow{-2}{*}{PESQ}          \\
\hline

\ac{HSI}       &\cellcolor[gray]{0.50}1.00           &\cellcolor[gray]{0.71}0.59           &\cellcolor[gray]{0.87}-0.26          &\cellcolor[gray]{0.99}0.02           &\cellcolor[gray]{0.77}0.47           &\cellcolor[gray]{0.65}0.71           &\cellcolor[gray]{0.77}0.47           \\
$-\mathcal{L}_\mathrm{WLM}$&\cellcolor[gray]{0.71}0.59           &\cellcolor[gray]{0.50}1.00           &\cellcolor[gray]{0.86}-0.29          &\cellcolor[gray]{0.96}-0.08          &\cellcolor[gray]{0.79}0.43           &\cellcolor[gray]{0.63}0.74           &\cellcolor[gray]{0.65}0.70           \\
$-\mathcal{L}_\mathrm{SNR}$&\cellcolor[gray]{0.87}-0.26          &\cellcolor[gray]{0.86}-0.29          &\cellcolor[gray]{0.50}1.00           &\cellcolor[gray]{0.89}0.22           &\cellcolor[gray]{0.92}-0.16          &\cellcolor[gray]{0.87}-0.26          &\cellcolor[gray]{0.85}-0.31          \\
SI-SNR         &\cellcolor[gray]{0.99}0.02           &\cellcolor[gray]{0.96}-0.08          &\cellcolor[gray]{0.89}0.22           &\cellcolor[gray]{0.50}1.00           &\cellcolor[gray]{0.96}0.08           &\cellcolor[gray]{0.90}0.19           &\cellcolor[gray]{0.97}-0.05          \\
HASPI          &\cellcolor[gray]{0.77}0.47           &\cellcolor[gray]{0.79}0.43           &\cellcolor[gray]{0.92}-0.16          &\cellcolor[gray]{0.96}0.08           &\cellcolor[gray]{0.50}1.00           &\cellcolor[gray]{0.69}0.62           &\cellcolor[gray]{0.86}0.27           \\
STOI           &\cellcolor[gray]{0.65}0.71           &\cellcolor[gray]{0.63}0.74           &\cellcolor[gray]{0.87}-0.26          &\cellcolor[gray]{0.90}0.19           &\cellcolor[gray]{0.69}0.62           &\cellcolor[gray]{0.50}1.00           &\cellcolor[gray]{0.68}0.64           \\
PESQ           &\cellcolor[gray]{0.77}0.47           &\cellcolor[gray]{0.65}0.70           &\cellcolor[gray]{0.85}-0.31          &\cellcolor[gray]{0.97}-0.05          &\cellcolor[gray]{0.86}0.27           &\cellcolor[gray]{0.68}0.64           &\cellcolor[gray]{0.50}1.00           \\

        \bottomrule
    \end{tabular}
    \end{footnotesize}
    }
    \label{table:correlations}
\end{table}

One would expect that a higher \ac{SNR} value (lower $\mathcal{L}_\mathrm{SNR}$) would indicate a higher intelligibility rating. \autoref{table:correlations} and the scatter plot in \autoref{fig:correlations} (top left) indicate the opposite; the signals with higher \ac{SNR} did not necessarily indicate that they would have higher intelligibility. This is likely to be because the signals in the CPC2 dataset were not originally optimised with an SNR loss and do not necessarily match the levels of the references.

In contrast, $\mathcal{L}_\mathrm{WLM}$ shows considerably stronger correlation with human intelligibility shown in \autoref{fig:correlations} (top right) and the other metrics compared with $\mathcal{L}_\mathrm{SNR}$. The particularly strong correlation of $\mathcal{L}_\mathrm{WLM}$ with \ac{PESQ} and \ac{STOI} in the \ac{CPC2} data is consistent with other datasets~\cite{closePerceivePredictSelfsupervised2023}. This strength speaks to the suitability of $\mathcal{L}_\mathrm{WLM}$ as a loss function to train the denoising component of a hearing aid system; training the denoising model to minimize this distance should by proxy improve the performance in terms of these metrics. 

Additionally, despite being the only metric considered which takes into account the severity of a listener's hearing loss i.e is specifically designed to predict intelligibility scores for hearing impaired people, \ac{HASPI} shows a weaker correlation with human intelligibility labels when compared with $\mathcal{L}_\mathrm{WLM}$ and \ac{STOI} and has a similar level of correlation compared to \ac{PESQ}, which is not aimed at assessing intelligibility. This brings into question the usefulness of \ac{HASPI} as a metric. Alternatively, it indicates that the severity of hearing loss does not actually factor massively into distribution of the human intelligibility labels, which is consistent with the findings in~\cite{close2023non}.  

\section{Hearing Aid System}

The hearing aid enhancement system trained in this work is an MC-Conv-TasNet-based system~\cite{tuTwoStageEndtoEndSystem}, which performed best in the \ac{CEC1}, consisting of two independently trained systems (one for each ear). Since modern hearing aids have multiple microphones, the system for each ear will take a multi-channel input, and the system for each ear will output a single channel.


\label{sec:system}
\begin{figure}[!ht]
    \centering
    \includegraphics[width=\columnwidth]{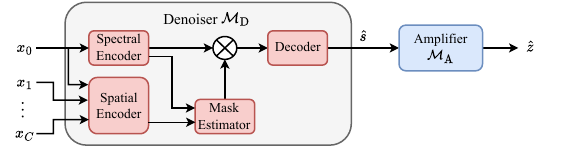}
    \caption{Workflow of the hearing aid system. For a $C$-channel signal $\mathbf{x}=\left[x_0,\dots,x_C\right]$, a reference channel $x_0$ is input to the spectral encoder, while the spatial encoder takes all $C$ channels as input.}
    \label{fig:system}
\end{figure}

This is a two-stage system as illustrated in \autoref{fig:system}, consisting of an MC-Conv-TasNet denoising block $\mathcal{M}_\mathrm{D}$ \cite{zhangEndtoendMultichannelTime2020}, followed by a \ac{FIR} filter block $\mathcal{M}_\mathrm{A}$  to amplify the signal according to a specific listener's audiogram \cite{tuOptimisingHearingAid2021}. During training, the output of $\mathcal{M}_\mathrm{A}$ is passed into a differentiable hearing loss simulation before the loss function is applied.

The overall system consists of $4$ neural networks: a denoiser $\mathcal{M}_\mathrm{D}$ for each stereo channel, and an amplifier $\mathcal{M}_\mathrm{A}$ for each stereo channel. For each ear, the denoiser $\mathcal{M}_\mathrm{D}$ is trained first, then the parameters are frozen for training of the amplifier $\mathcal{M}_\mathrm{A}$.

\subsection{\texorpdfstring{Denoiser $\mathcal{M}_D$}{Denoiser}}

The structure of MC-Conv-TasNet is given in \autoref{fig:system}. This model consists of a spectral encoder, a spatial encoder, a mask estimation network, and a decoder. The spectral encoder is a $1$D convolutional network which takes a single reference channel from the multi-channel mixture and which has a kernel size $L$. The spatial encoder takes a multi-channel signal $\mathbf{x}$ as input and operates as a $2$D convolutional network over all channels; if $C$ denotes the number of microphone channels, the kernel size of the spatial encoder is $C\times L$ so that the number of frames of the spatial encoder is the same as the number of frames of the spectral encoder. The encoder outputs are concatenated as they are input to the mask estimator, which is a temporal convolutional network \cite{leaTemporalConvolution2016}. The mask is multiplied element-wise with the spectral encoder output to give the features of the estimated source. Finally, a decoder reconstructs a single-channel waveform as an estimate for the target.

The configuration of $\mathcal{M}_D$ is exactly as described in \cite{tuTwoStageEndtoEndSystem}, where the spectral and spatial encoders use $256$ and $128$ filters, respectively, both with a frame length of $L=20$ samples. The $1\times1$ bottleneck convolutional block uses $256$ channels and the convolutional blocks use $512$ channels. There are $6$ convolutional blocks, which have a kernel size of $3$ and dilation factors of $1,2,\dots,32$ and are repeated $4$ times. The input to $\mathcal{M}_D$ is the time domain multi-channel noisy signal $\mathbf{x}$ and the output is the denoised single channel audio $\hat{s}$. 

\subsection{\texorpdfstring{Amplifier $\mathcal{M}_A$}{Amplifier}}
The \ac{FIR} filter amplification module \cite{tuOptimisingHearingAid2021}, $\mathcal{M}_A$, is a simple, linear amplification model designed to compensate for hearing loss in each ear without introducing distortions and artefacts.

As with the denoiser, two independent amplification models are trained (one for each ear), taking the $22.05$kHz, single channel output of the denoising module $\mathcal{M}_D$. The result is upsampled to $44.1$kHz, clipped to the range $[-1,1]$, and fed into the differentiable hearing loss simulator $\mathcal{M}_{\mathrm{HL}}$. Upsampling is required as the hearing loss simulator used operates at $44.1$kHz.





\section{Experiment Setup}
\label{sec:experiment}

\subsection{Clarity Enhancement Challenge Datasets}

Two datasets are used for training the system: the \ac{CEC1} dataset, for which the baseline system was designed, and the \ac{CEC2} dataset, which is a more challenging dataset for the same task. Both of these datasets consist of a train set containing $6000$ samples, a validation set of $2500$ samples, and an evaluation set of $1500$ samples. Each sample is a $6$-channel, noisy, reverberant mixture $\mathbf{x}$, with a target, anechoic speech signal $\mathbf{s}=\left[s_\mathrm{l}, s_\mathrm{r}\right]$ for each ear (left and right).

For \ac{CEC1}, there is only one interfering source in the noisy mixture. In half of the samples, this is a second speech signal sourced from \cite{demirsahinMultispeakerCorpora2020}, and in the other half, the interfering signals are domestic noise sourced from \cite{fontFreesound2013}. For \ac{CEC2}, there are $2$ or $3$ interfering sources in each noisy mixture, which could be competing speakers, music, or domestic noise. All data has a sampling frequency of $44.1$kHz.

\subsection{Training Setup}
\label{ssec:TrainingSetup}

In this task, maintaining the signal level is important for the downstream amplification, which is required for hearing aid users. With this in mind, we propose a joint loss function \begin{equation}
    \label{eqn:joint-loss}
    \mathcal{L}_\mathrm{Joint}\left( s, \hat{s}\right) = \mathcal{L}_\mathrm{SNR}\left( s, \hat{s}\right) + \mathcal{L}_\mathrm{WLM}\left( s, \hat{s}\right),
\end{equation}
where $\mathcal{L}_\mathrm{SNR}$ is used to maintain the signal level of the reference.

In early experiments, it was found that $\mathcal{L}_\mathrm{SNR}$ and $\mathcal{L}_\mathrm{WLM}$ optimise with different learning rates, with $\mathcal{L}_\mathrm{SNR}$ preferring a greater learning rate. Two different training strategies are used, which vary the learning rate over the training process.

All modules were trained using the Adam optimiser \cite{kingmaAdamMethodStochastic2017}. The learning rate is initially set to $10^{-3}$, and gradient clipping is applied with a maximum L$2$-norm of $5$.
Three training settings were used:
\begin{enumerate}
    \item \emph{Baseline} - The denoiser $\mathcal{M}_D$ was trained for $200$ epochs using the SNR loss function $\mathcal{L}_\mathrm{SNR}$  (\ref{eqn:SNR-loss}).
    \item \emph{WavLM Encoder loss fine-tuning} - $\mathcal{M}_D$ is trained for $100$ epochs using $\mathcal{L}_\mathrm{SNR}$ (\ref{eqn:SNR-loss}) followed by $100$ epochs of training using $\mathcal{L}_\mathrm{Joint}$ (\ref{eqn:joint-loss}). During the fine-tuning stage, the learning rate is set to $10^{-4}$.
    \item \emph{Joint Loss with learning rate scheduling} - $\mathcal{M}_D$ is trained using the joint loss function $\mathcal{L}_\mathrm{Joint}$ (\ref{eqn:joint-loss}) for $200$ epochs. A scheduled learning rate change inspired by~\cite{transformer} is used, with the learning rate linearly increasing for the first few training steps, and then with a learning rate decay over the remaining epochs.  
\end{enumerate}


Models for each (left and right) ear are trained separately; both sides take all $6$ channels of input (downsampled to $22.05$kHz), and the single corresponding left/right channel of the target audio is as the reference.

\section{Results}
\label{sec:results}

\autoref{tab:den_results} shows the evaluation metrics for the output of the denoiser $\mathcal{M}_\mathrm{D}$ for the baseline $\mathcal{L}_\mathrm{SNR}$ and the proposed $\mathcal{L}_\mathrm{WLM}$ for both the \ac{CEC1} and \ac{CEC2} test sets. On \ac{CEC1}, the system trained with the proposed loss function using a scheduled learning rate shows a significant improvement on \ac{HASPI}, \ac{STOI}, \ac{PESQ} and \ac{SI-SNR} scores over the $\mathcal{L}_\mathrm{SNR}$ baseline, while the proposed fine-tuning approach performs similarly to the baseline.

On the more challenging \ac{CEC2} dataset, there is an improvement in \ac{HASPI} and \ac{STOI} for the proposed loss function using a scheduled learning rate, but the fine-tuning approach performs worse than the baseline.  All systems give slightly lower \ac{PESQ} scores on this dataset versus the noisy input; this is interesting given the strong correlation between these two metrics shown in \autoref{table:correlations}. It should, however, be noted that the main task of CEC is increasing intelligibility.

\begin{table}[!h]
    \centering
    \caption{Performance of the denoiser $\mathcal{M}_\mathrm{D}$.}
    \resizebox{\columnwidth}{!}{
    \begin{tabular}{cl|ccccc}
         \multirow{2}{*}{Dataset}                  &\multirow{2}{*}{Model}                         &\multirow{2}{*}{\ac{HASPI}}     &\multirow{2}{*}{\ac{STOI}}      &\multirow{2}{*}{\ac{PESQ}}      &$\Delta$&$\Delta$\\
                           &{}                        &{}    &{}     &{}     &SI-SNR&fwSNR\\
\hline
\multirow{3}{*}{\rotatebox[origin=c]{90}{CEC1}}                         &$\mathcal{L}_{\mathrm{SNR}}$&0.90      &0.76      &1.19      &7.92      &\textbf{3.72}      \\
                         &$\mathcal{L}_{\mathrm{WLM}}$, FT&0.90      &0.76      &1.20      &8.03      &3.42      \\
                         &$\mathcal{L}_{\mathrm{WLM}}$, Sched.&\textbf{0.91}      &\textbf{0.78}      &\textbf{1.23}      &\textbf{8.45}      &1.65      \\
\hline
\multirow{3}{*}{\rotatebox[origin=c]{90}{CEC2}}                         &$\mathcal{L}_{\mathrm{SNR}}$&0.72      &0.67      &1.11      &10.09     &0.60      \\
                         &$\mathcal{L}_{\mathrm{WLM}}$, FT&0.72      &0.66      &1.10      &9.98      &0.57      \\
                         &$\mathcal{L}_{\mathrm{WLM}}$, Sched.&\textbf{0.75}      &\textbf{0.68}      &\textbf{1.12}      &\textbf{10.33}     &\textbf{0.78}      \\
    \end{tabular}
    }
    \label{tab:den_results}
\end{table}

On \ac{CEC1}, the \ac{HASPI} improvement is only small, though the scores for all systems are very high for this metric. On \ac{CEC2}, the improvement over the baseline is larger, and on this more challenging dataset, there is a perceptual improvement in these signals for listeners with hearing impairment.



\section{Conclusion}
\label{sec:conclusion}

In this work, it is shown that encoder representations obtained from WavLM can effectively capture the intelligibility of a speech signal. This is shown by implementing a simple distance function between representations of noisy and clean speech signals and correlating these distances with human intelligibility labels and perceptually motivated metrics for both intelligibility and quality. Further, this work evaluates the use of this distance in a loss function for training the denoising network of a hearing aid system. This new training setting shows improved performance over a system trained with a purely signal-based loss function, with improvements on \ac{HASPI}, \ac{STOI}, \ac{PESQ}, and \ac{SI-SNR}.



\bibliographystyle{IEEEbib-abbrev}
\bibliography{ref}

\end{document}